\documentclass[a4paper,prb]{revtex4}
\usepackage{amsmath}
\usepackage{graphicx}
\usepackage{color}
\usepackage{subfigure}
\usepackage{hyperref}
\begin{document}
\title{Water Propagation in the Porous Media, Self-Organized Criticality\\
 and Ising Model}
\author {M. N. Najafi}
\email{morteza.nattagh@gmail.com}
\affiliation{Department of Physics, University of Mohaghegh Ardabili, P.O. Box 179, Ardabil, Iran}
\author {M. Ghaedi}
\affiliation{Chemical and Petroleum Engineering Department, Sharif University of Technology, P.O. Box 11365-9465, Tehran, Iran}
\begin{abstract}
In this paper we propose the Ising model to study the propagation of water in 2 dimensional (2D) petroleum reservoir in which each bond between its pores has the probability $p$ of being activated. We analyze the water movement pattern in porous media described by Darcy equations by focusing on its geometrical objects. Using Schramm-Loewner evolution (SLE) technique we numerically show that at $p=p_c\simeq 0.59$, this model lies within the Ising universality class with the diffusivity parameter $\kappa=3$ and the fractal dimension $D_f=\frac{11}{8}$. We introduce a self-organized critical model in which the water movement is modeled by a chain of topplings taking place when the amount of water exceeds the critical value and numerically show that it coincides with the numerical reservoir simulation. For this model, the behaviors of distribution functions of the geometrical quantities and the Green function are investigated in terms of $p$. We show that percolation probability has a maximum around $p=0.68$, in contrast to common belief.
\end{abstract}
\maketitle
\section{Introduction}
In petroleum production, the reservoir simulation as a method of evaluation of different production scenarios, determination of the amount of economic production and verification of the impact of uncertainties in key reservoir parameters, plays a vital role.  A lot of attention has been paid to the modeling of fluid motion in porous media and different approaches have been used such as numerical simulations \cite{Ertekin, Aziz1, Watts, Allen}, pore network modeling \cite{Al-Dhahli, Blunt1, Blunt2} and percolation theory for the porous systems\cite{Sadeghnejad,Masihi,Sahimi}. One of the most important challenges in this area is the petroleum production optimization. To optimize the petroleum production from a reservoir, an enhanced oil recovery (EOR) method can be utilized. A common method of oil recovery is by displacement in which the water or the gas is injected into a well\cite{injection1,injection2}. The injected fluid moves from the injection well towards the production well. Thus the percolation theory may be employed to find the percent of the injected fluid which is percolated to the production well\cite{Sahimi,percolationtheory}. Generally the study of reservoir connectivity is often important for EOR processes and well placement. For instance, during water flooding, sweep efficiency of the process is controlled by inter-well connectivity. Percolation theory via simulating sandbody connectivity behaviors can be used to estimate inter-well connectivity by determining the percolation probability\cite{percolationtheory,Masihi}. Due to its considerable effect on the amount of oil recovery, the movement pattern of the injected fluid is directly addressed in optimization problem\cite{percolationtheory}. In the other hand, the complex nature of hydrocarbon bearing formation of such media's has a considerable effect on the fluid motion and its pattern in the porous media. Knowing that these media's may be divided into permeable and impermeable parts, one can assume that the fluid movement takes place only in the permeable parts of the reservoir. Examples of the such reservoirs are sandstone reservoirs with shale, naturally fractured reservoirs and channel reservoirs\cite{reservoirsEX}. The movement of water in such media's is usually considered as passive tracer transport which means that there is no interaction between the water and the containing reservoir fluid. The injected water in a pore is static, until it reaches a certain saturation known as "critical water saturation" after which the water flows freely to the neighboring pores\cite{Blunt1}. We name this procedure as "water toppling". Despite of much theoretical attempts carried out towards understanding this dynamics, an exact mechanism, resulting from this "water toppling" is missing in the literature yet.\\
Conventionally to predict the behaviors of reservoirs, a detailed geological and flow model is built and after upscaling, it can be used for simulating the flow dynamics. For review see \cite{Aziz1,Fanchi1,Fanchi2}. The mentioned procedure is computationally expensive and time-consuming. Thus sometimes the number of constructed realizations are reduced which may not be reliable. Despite of the recent advances in enhancing the simulation speed, such as upscaling and streamline simulation, this problem has still remained\cite{Blunt3, Datta, Thiele}. The aim of this paper is to investigate the movement of the injected water in a reservoir via focusing on the geometrical objects. We first investigate the numerical reservoir simulation of water propagation in 2D porous media described by Darcy equations and a long with this model, we introduce a self organized critical model of this propagation (SOC model)\footnote{Usually a narrow or a stratified reservoir can be regarded as a 2D reservoir.}. The main feature of the SOC model is the notion of self-organized criticality (SOC) introduced by Bak, Tang and Wiesenfeld (BTW) \cite{BTW} as non-equilibrium systems which show robust critical behavior without fine tuning of any parameter. Unlike the ordinary critical systems, these systems may be open and dissipative and energy input is necessary to offset the dissipation. As a prototype, BTW introduced a simple model, i.e. the Abelian sandpile model (ASM) on which many theoretical and numerical investigations have been done\cite{Dhar2,Dhar3,najafi}. The Gutenberg-Richter law in Earthquake, Scheidegger's model of river basins, Takayasu's aggregation model and the voter model are the examples of the models which are directly related to ASM\cite{Dhar3}. The mechanism of the water toppling is reminiscent of the pile toppling in the sandpile models. In the present paper we analyze numerically the domainwalls which are defined as the separator of the so called "toppled" and "untoppled" pores of the media by means of Schramm-Loewner evolution (SLE) technique.\\
SLE theory is a candidate to classify 2D critical statistical models in one parameter classes\cite{Schramm}. According to this theory one can describe the geometrical objects (which may be interfaces) of a 2D critical model via a growth process. The essential belding block of SLE is the conformal symmetry of the probability measures of the model in hand. The connection between SLE and conformal field theory (CFT) shows itself in a simple relation between the the diffusivity parameter $\kappa$ in SLE and the central charge $c$ in CFT, proposed by M. Bauer et. al. \cite{BauBer}. Using SLE technique, we show that these models belong to the universality class of the Ising model with the diffusivity parameter $\kappa=3$ and CFT with the central charge $c=\frac{1}{2}$. We also address the problem of the probability of water percolation in a reservoir. We see that the percolation probability (as well as the correlation length and the Green function) has a maximum at an especial amount of $p$. In contrast to common belief, we find that it takes place at $p\simeq 0.68$.\\
The paper has been organized as follows: In Sec [\ref{SOC}] we have introduced two models, namely reservoir flow model and SOC model, to investigate the water propagation in the porous media and in Sec [\ref{connection}], by means of Schramm-Loewner evolution (SLE) technique, we numerically show that these two models are in the same universality class. Further numerical results for the SOC model are presented in Sec [\ref{num}].

\section{Models of Water Movement in the Porous Media}\label{SOC}
In this section we briefly introduce the well-known reservoir flow model and the SOC model to investigate the dynamics of water in the porous media and the connection between these models will be studied in the next section. Porous media is a material comprising of solid matrix with connected or unconnected pores. It is mainly described by its porosity which is the ratio of the void space to the total volume of the medium, and its permeability which is a measure of the ability of a porous media to allow fluids to pass through. For each model, we consider a $L\times L$ square lattice in which two neighboring sites are connected with the probability $p$ and are disconnected with the probability $1-p$. The resulting lattice is a 2D porous media and can be analyzed via the percolation theory. For $p\geq p_c=0.5927$ (for the square lattice) there is a finite probability of having a percolated cluster, i. e. a cluster of the linear size of the system.\\
\\
\textbf{Reservoir Flow Model}\\
 According to reservoir flow model, to handle this dynamics one may use two laws, namely mass conservation law and Darcy's law of velocity. The resulting non-linear differential equation is as follows:
\begin{equation}
\partial_t \left( \phi\rho_{\alpha}S_{\alpha}\right) +\nabla .\left( \frac{-Kk_{r\alpha}\rho_{\alpha}}{\mu_{\alpha}}\left( \nabla p_{\alpha}-\rho_{\alpha}g\nabla d\right) \right) =q_{\alpha}
\label{NewModel}
\end{equation}
in which $\alpha$ stands for the fluid phases (oil or water), $\phi$ is the system porosity, $S$ is the saturation, $q$ is the source or sink term, $\rho$ is the fluid density, $g$ is the gravitational acceleration of earth, $d$ is the vertical coordinate (measured from depth)\footnote{The gradient of $d$ in our analysis is zero, because we assume 2D lattice parallel to the earth surface}, $K$ is the absolute permeability tensor, $k_r$ is the phase mobility and $\mu$ is viscosity.\\
For a three dimensional isotropic reservoir, $K$ is a diagonal $3\times 3$ matrix with the identical elements, i.e. $K=\text{diag}(k,k,k)$. Writing this equation for each phase will result in two equations. Two auxiliary equations also are used in reservoir simulations to find all unknown quantities:
\begin{equation}
\begin{split}
&S_o+S_w=1\\
&p_o+p_w=p_c(S_w)
\end{split}
\label{NewModel2}
\end{equation}
where $p_c$ is the capillary pressure. It is worth mentioning that the water transfer occurs only trough the activated bonds and this brings the occupancy probability $p$ to the story. One may initially begin with an empty lattice and inject water to a random site and let the system to be relaxed by means of Eq [{\ref{NewModel}]. After relaxation, the water is injected to another random site again and the process continues. This leads the system to have non-zero average water saturation. Showing $n$ as the water injection step, we can plot this average in terms of $n$ which has been shown in FIG [\ref{steady}]. This figure shows that, in some $n$, the system reaches a steady state after which the pressure is statistically constant. We analyze the model at the steady states. After each relaxation, we can calculate the change of the amount of water in each site form its initial value and label this site as "toppled" if its amount is larger than some value $\delta$. We have found that our analysis is nearly independent of $\delta$. Thus after each relaxation (which we name it as "water toppling"), we will have some "toppled" and "untoppled" sites and we can define some non-intersecting loops in the lattice which separates "toppled" and "untoppled" sites. The set of all toppled sites is called an avalanche.
\begin{figure}
\centerline{\includegraphics[scale=.40]{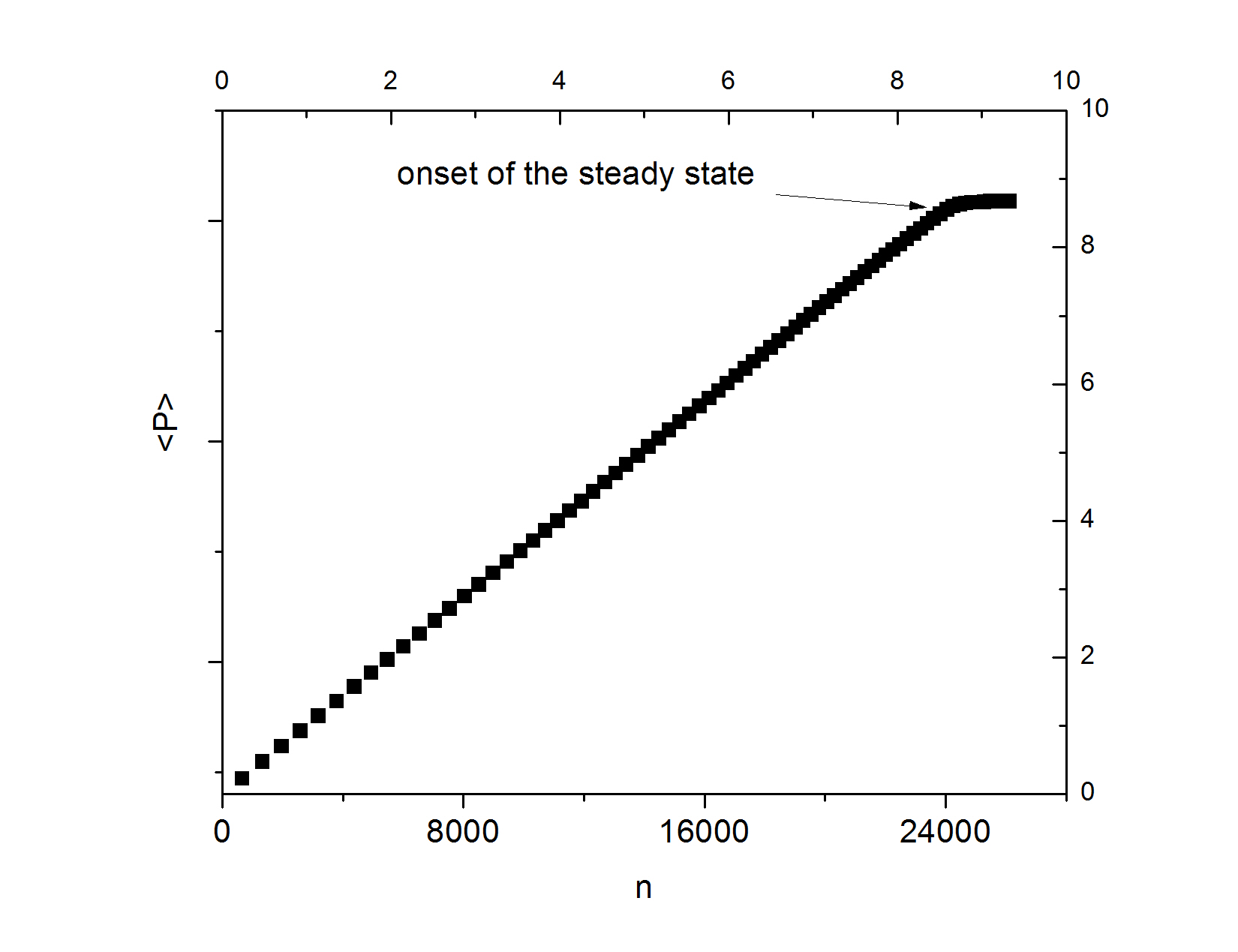}}
\caption{Average value of water saturation versus the injection step.}
\label{steady}
\end{figure}
 \\
 \\
\textbf{Self Organized Critical Model}\\
Reservoir fluid flow equations are difficult to handle, because they consist of couple of non-linear partial differential equations and the CPU time for the numerical solutions rapidly grows with the system size. It makes this model inefficient for statistical analysis. In this section we introduce a simple model to investigate the dynamics of the reservoir. \\
Lets assign to each site an amount of water ($h_i$) lower than a critical value above which the water overflows to the neighboring sites. We consider the quantized water transfer during each relaxation process. When we inject the water at a random site, $h_{i}\rightarrow{h_{i}+1}$ the amount of its water may exceed the critical value and the water "topples" to the sites which are connected to it. We consider no dissipation, i. e. if the original site has $z$ connected neighbors, during the toppling, the amount of water of the original site decreases by $z$ and the connected neighboring sites increase by one. As a result, the neighboring sites may become unstable and topple and a chain of topplings may happen in the system. In the boundary sites, the toppeling causes one or two sands leave the system. This process continues until the system reaches to a stable configuration. The process is then repeated. When the system reaches a stable configuration, each site of the lattice have one of two possible states, toppled or untoppled. 
\begin{figure}
\centerline{\includegraphics[scale=.40]{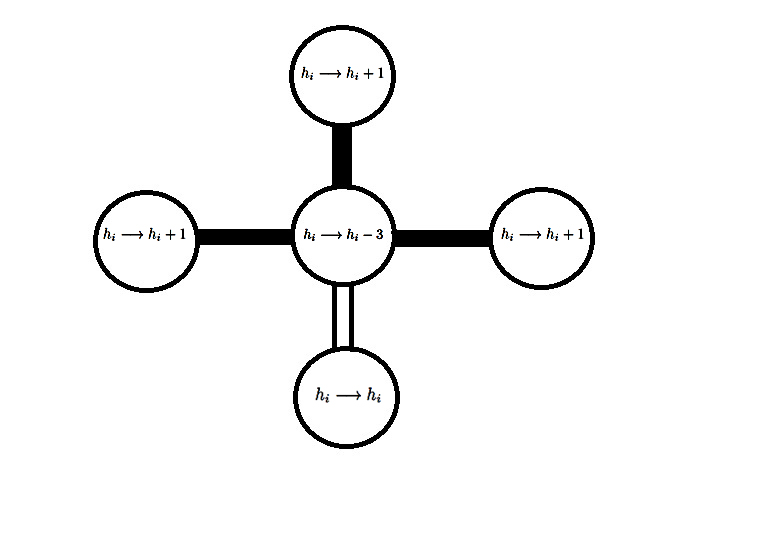}}
\caption{The toppling rule for a typical site which have 3 connected neighbors.}
\label{samplea}
\end{figure}
\begin{figure}
\centerline{\includegraphics[scale=.40]{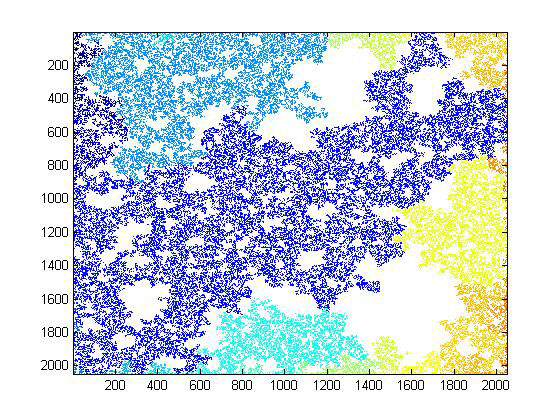}}
\caption{A sample of a percolation cluster for $p=p_c$.}
\label{sampleb}
\end{figure}
When $p=1$, one retrieves the BTW \cite{BTW} model. In this model one assigns to each site a height variable (the number of grains in the site) taking its values from the set $\lbrace{1, 2, 3, 4}\rbrace$ in such a way that each configuration of the resulting "sand pile" is given by the set $\lbrace{h_{i}}\rbrace$. After adding a grain to a random site $i$, if the resulting height becomes more than 4, the site topples and loses 4 grains, each of which is transferred to one of four neighbours of the original site. The dynamics of the system is like above. The movement in the space of stable configurations lead the system to fall into a subset of configurations after a finite steps, named as the "recurrent states" in which a configuration may happen regularly and the system reaches the steady state. In this state, the recurrent states all occur with the same probability. It has been shown that the total number of recurrent states is det$\Delta$ where $\Delta$ is the discrete Laplacian. For details see \cite{Dhar3}. This model can be generalized to other lattice geometries and to off critical set up\cite{najafi}. It has been shown that in the continuum limit, this model corresponds to $c=-2$ conformal field theory (with the ghost action $S=\int{d^{2}z}(\partial\theta\bar{\partial}\bar{\theta})$ in which $\theta$ and $\bar{\theta}$ are Grassmann variables) and the frontier of its avalanches are loop-erased random walks (LERW) which are $\text{SLE}_{\kappa=2}$\cite{najafi}.\\

The same features in the SOC model of water movement in the porous media is observed. After a finite number of water injection (of order $0.6L_x\times L_y$ for a system of size $L_x\times L_y$) the system reaches a steady state in which the average amount of water ($\bar{h_i}$) becomes statistically constant. In FIG [\ref{samplea}] we have shown schematically the toppling rule for a site with 3 connected neighboring sites. FIG [\ref{sampleb}] shows a typical graph of a porous system for $p=0.59$. Only the clusters which are connected to the boundary of the lattice have been considered. 

\section{Connection between two models; Ising Universality class}\label{connection}

Now we investigate numerically the equivalence of two models mentioned above. For the Darcy model simulation, we have considered $200\times 200$ square lattice and over $6\times 10^3$ samples have been generated\footnote{An open source MATLAB toolbox has been described in ref\cite{Lie}. The MATLAB Reservoir Simulation Toolbox (MRST) is developed by SINTEF Applied Mathematics. For more information visit http://www.sintef.no/Projectweb/MRST/, along with the references therein.}.  Equal viscosity of $1cp$ is considered for both oil and injected water. Zero capillary pressure is considered and the relative permeability for water and oil are calculated by the following simple relations:
\begin{equation}
\begin{split}
&k_{rw}=S_w-0.2\\
&k_{ro}=1-S_w
\end{split}
\label{conditions}
\end{equation}
In this simulation, 100 md absolute permeability is assigned to occupied sites and zero to non-occupied ones. The initial water saturations are randomly distributed among the grid cells such that the saturations are less than the critical saturation which is 0.2.\\
For the SOC model, $512\times 512$ square lattice has been considered and over $10^5$ samples have been generated. For each model we have extracted the domainwalls which are exterior frontiers of the avalanches, shown in FIG [\ref{DW-samplesa}] (for the Darcy model) and FIG [\ref{DW-samplesb}] (for the SOC model) (for $p=0.5927$) as an example. Two statistical quantities have been calculated: fractal dimension and the diffusivity parameter ($\kappa$) in the SLE theory. 
\begin{figure}
\centerline{\includegraphics[scale=.40]{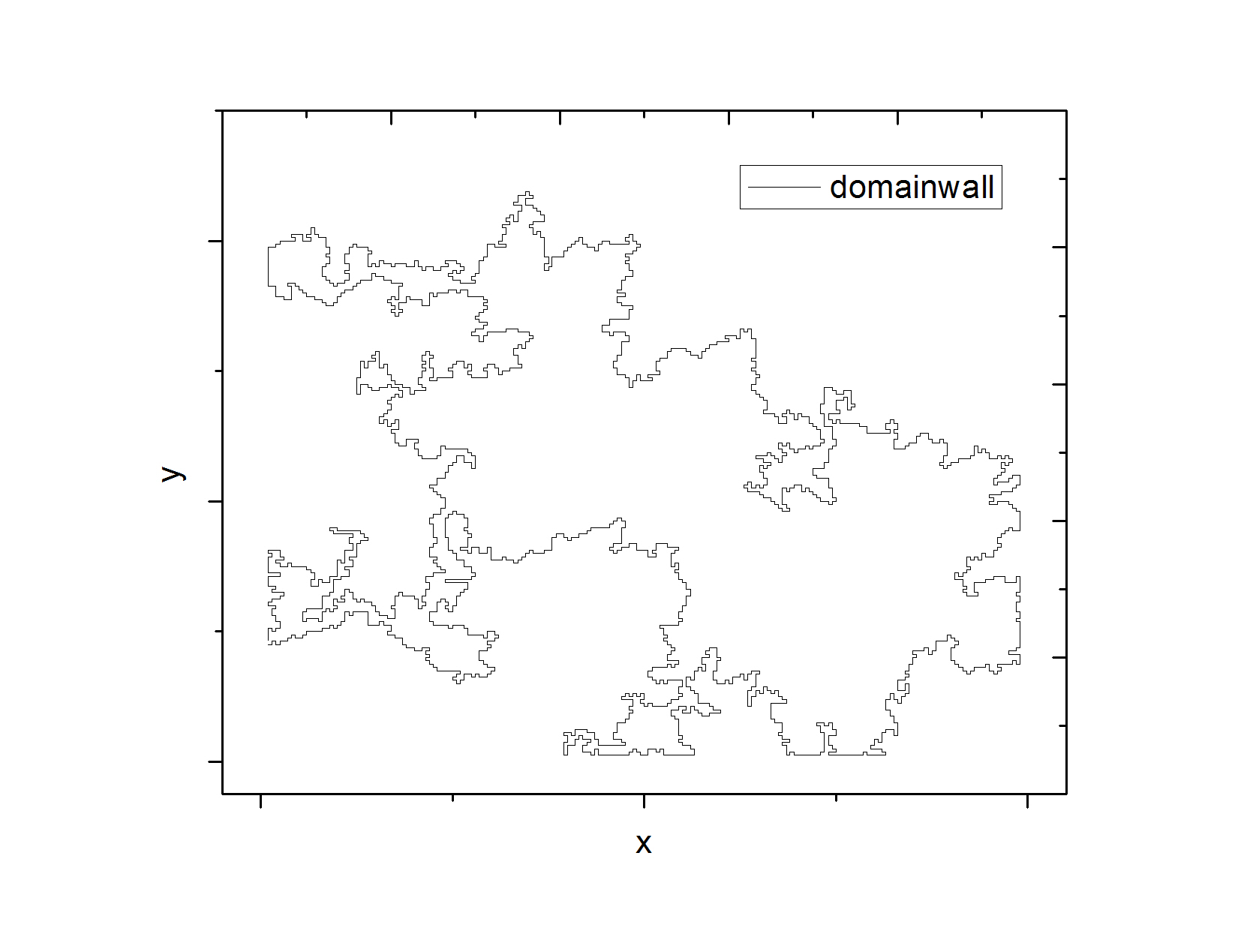}}
\caption{A domain wall sample of the propagated water in the porous medium.}
\label{DW-samplesa}
\end{figure}
\begin{figure}
\centerline{\includegraphics[scale=.40]{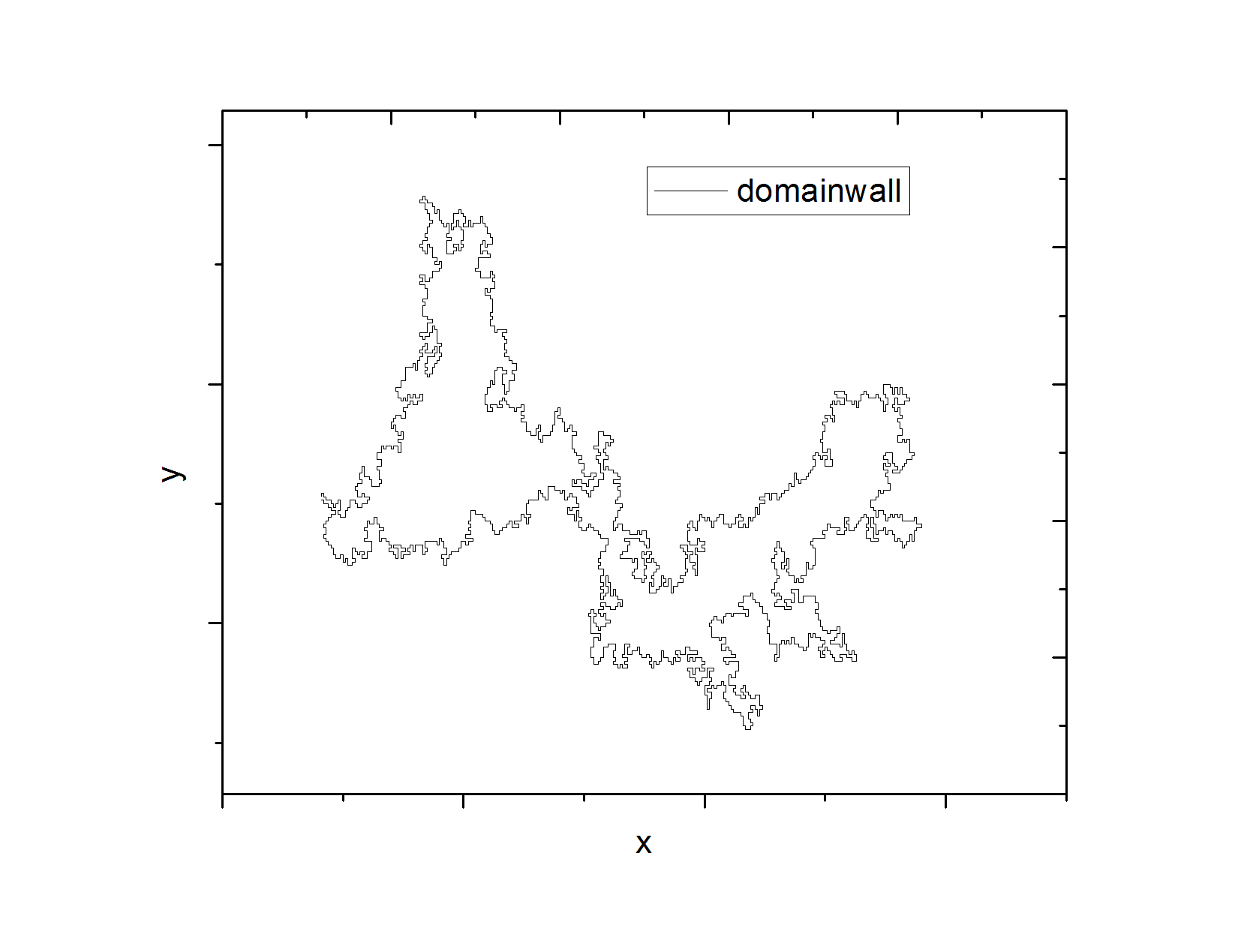}}
\caption{A domain wall sample of the propagated water in the SOC model.}
\label{DW-samplesb}
\end{figure}
\begin{figure}
\centerline{\includegraphics[scale=.40]{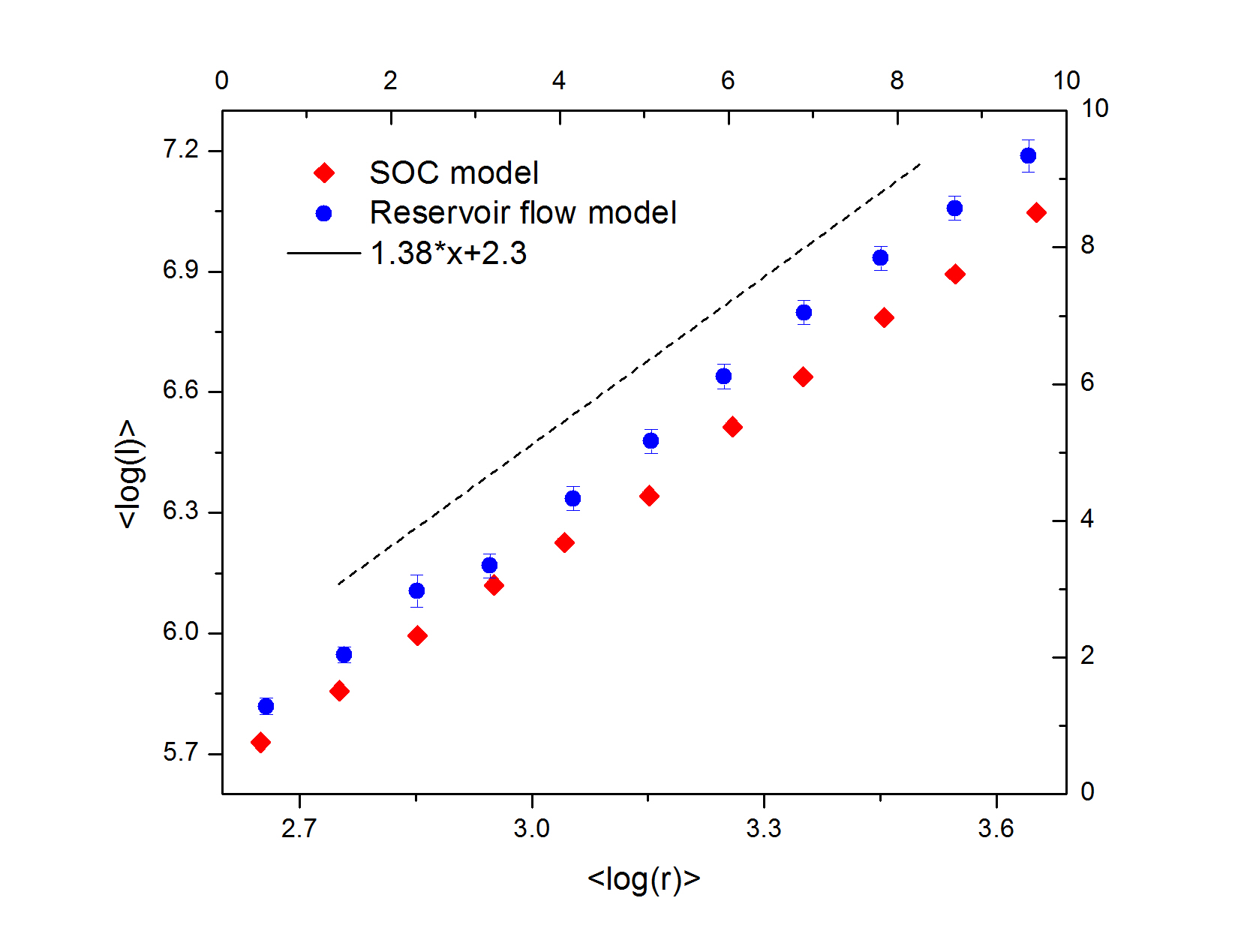}}
\caption{The plot of $\left\langle \log(l)\right\rangle$ versus $\left\langle \log(r)\right\rangle $ for the reservoir flow model and SOC model. }
\label{FDa}
\end{figure}
\begin{figure}
\centerline{\includegraphics[scale=.40]{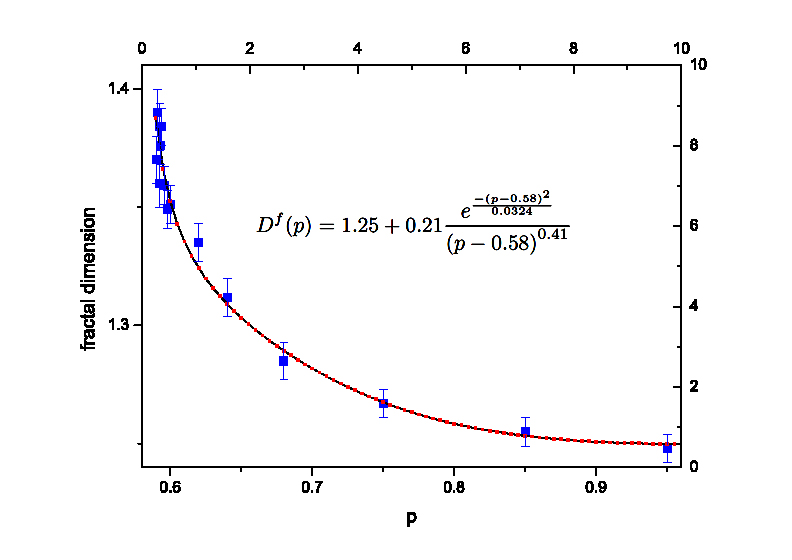}}
\caption{The fractal dimension of the curves in the SOC model versus $p$.}
\label{FDb}
\end{figure}
\\
\\
\textbf{Fractal Dimension}\\
The fractal dimension ($D_f$) of the loops is defined by the relation $\left\langle \log(l)\right\rangle \sim D_f \left\langle \log (r)\right\rangle $ in which $l$ is the loop length and $r$ is the loop gyration radius. FIG[\ref{FDa}] shows that the calculated $D_f$ for two models at $p=p_c$ are identical. The main point is that this value is $1.38\simeq\frac{11}{8}$ corresponding to the Ising model. We also found that the samples with $p\geq p_c$ are self-similar and show critical behaviors. In FIG [\ref{FDb}] we have presented the fractal dimension of the curves in the SOC model in terms of $p$. This graph is fitted by $D_f=D_f^0+a\frac{e^{- \frac{(p-p_c)^2}{\alpha}}}{(p-p_c)^\beta} $ with $D_f^0\simeq1.25 , a\simeq 0.21, \alpha\simeq 0.034 ,\beta\simeq 0.41$. The main feature of this figure is the singular behavior which shows itself in the rapid oscillations of fractal dimension near $p_c$. This change of behavior at $p_c$ is due to the singular behavior of percolation theory near $p_c$. The fractal dimension approaches to $1.25$ at $p=1$ corresponding to the BTW model.
\begin{figure}
\centerline{\includegraphics[scale=.40]{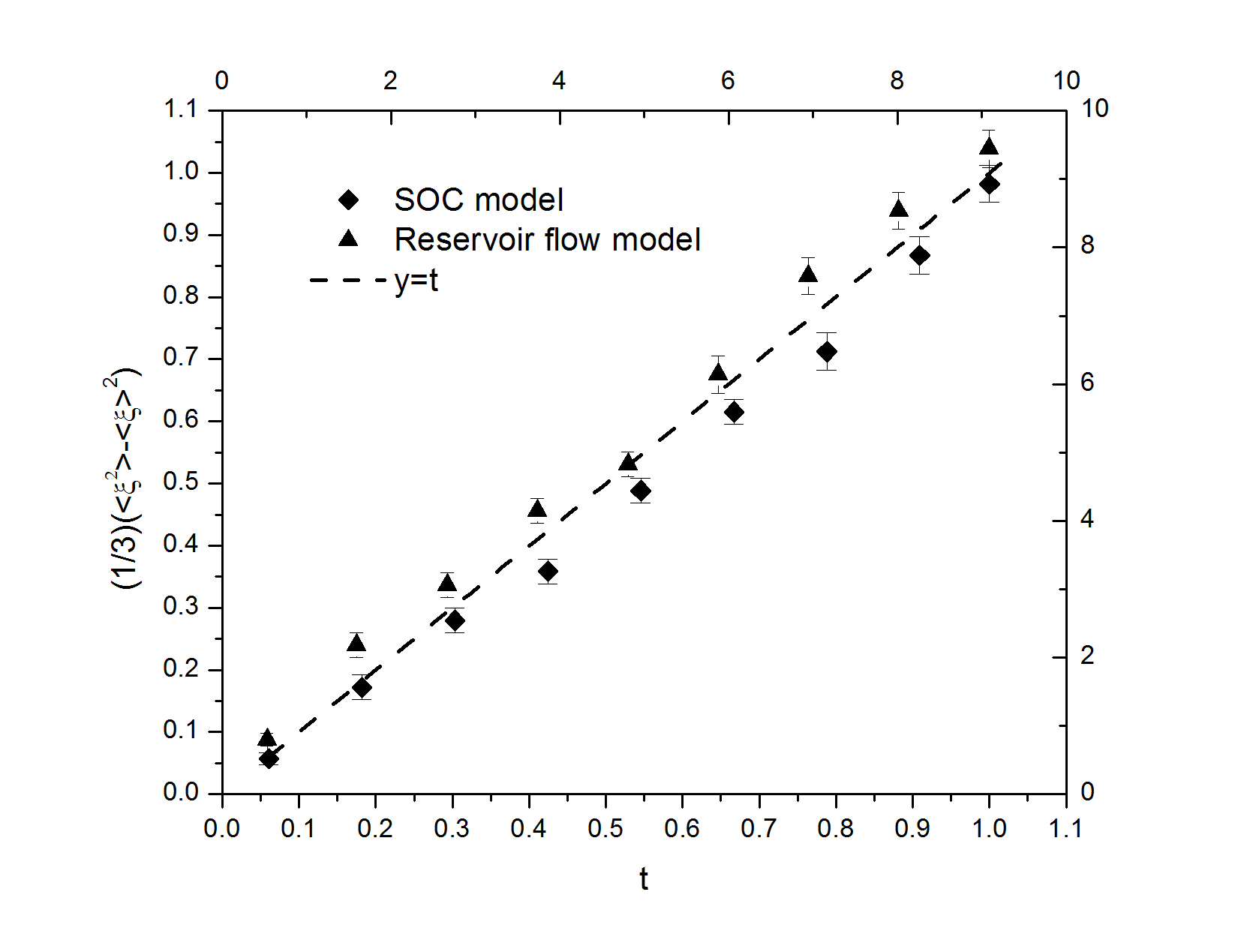}}
\caption{$\left( \langle B_t^{2}\rangle-\langle B_t\rangle^{2}\right)$ versus $t$ for the case $p=p_c$.}
\label{SLE}
\end{figure}
\begin{figure}
\centerline{\includegraphics[scale=.40]{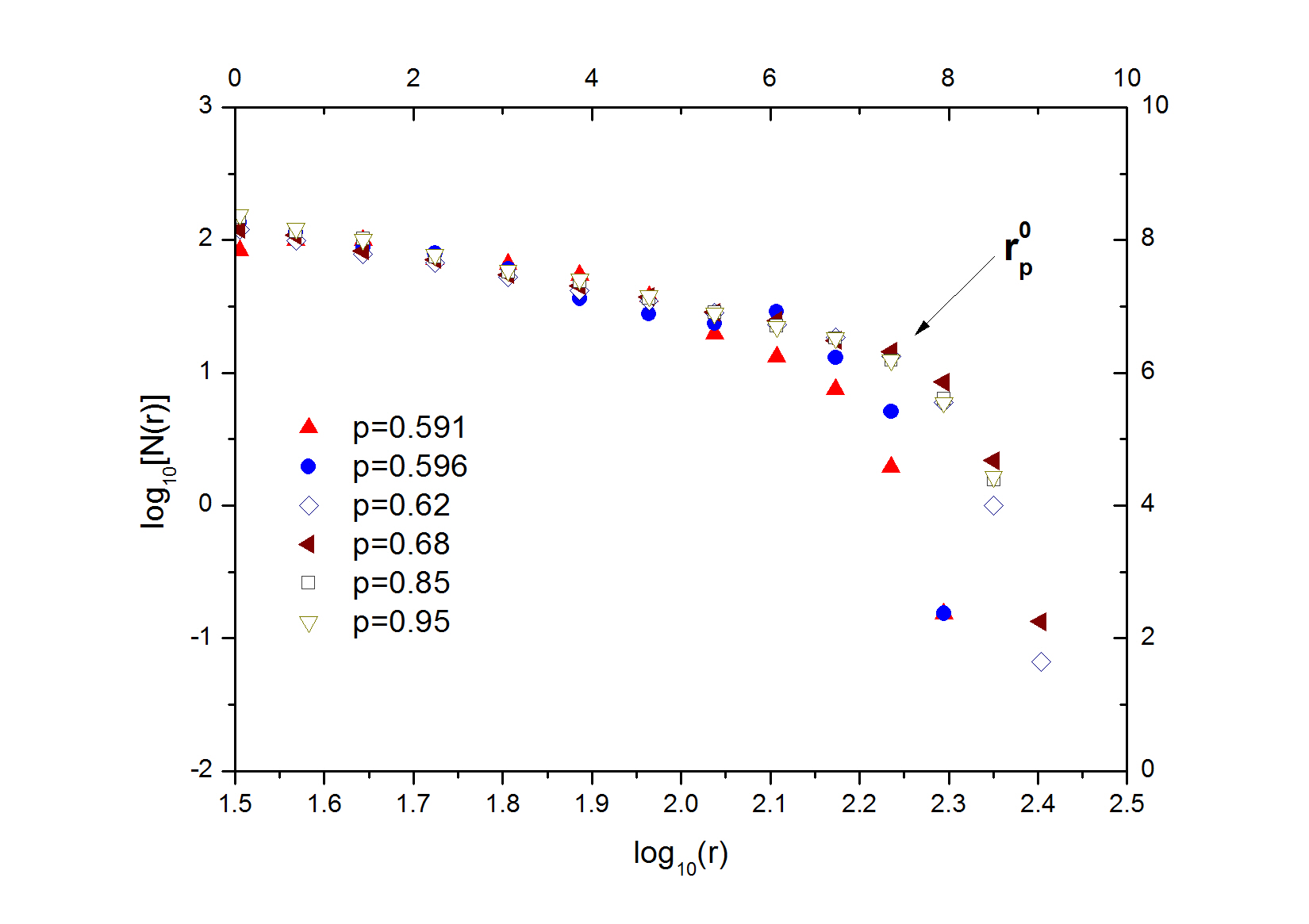}}
\caption{The distribution function of gyration radius for various rates of $p$.}
\label{gyrationa}
\end{figure}
\begin{figure}
\centerline{\includegraphics[scale=.40]{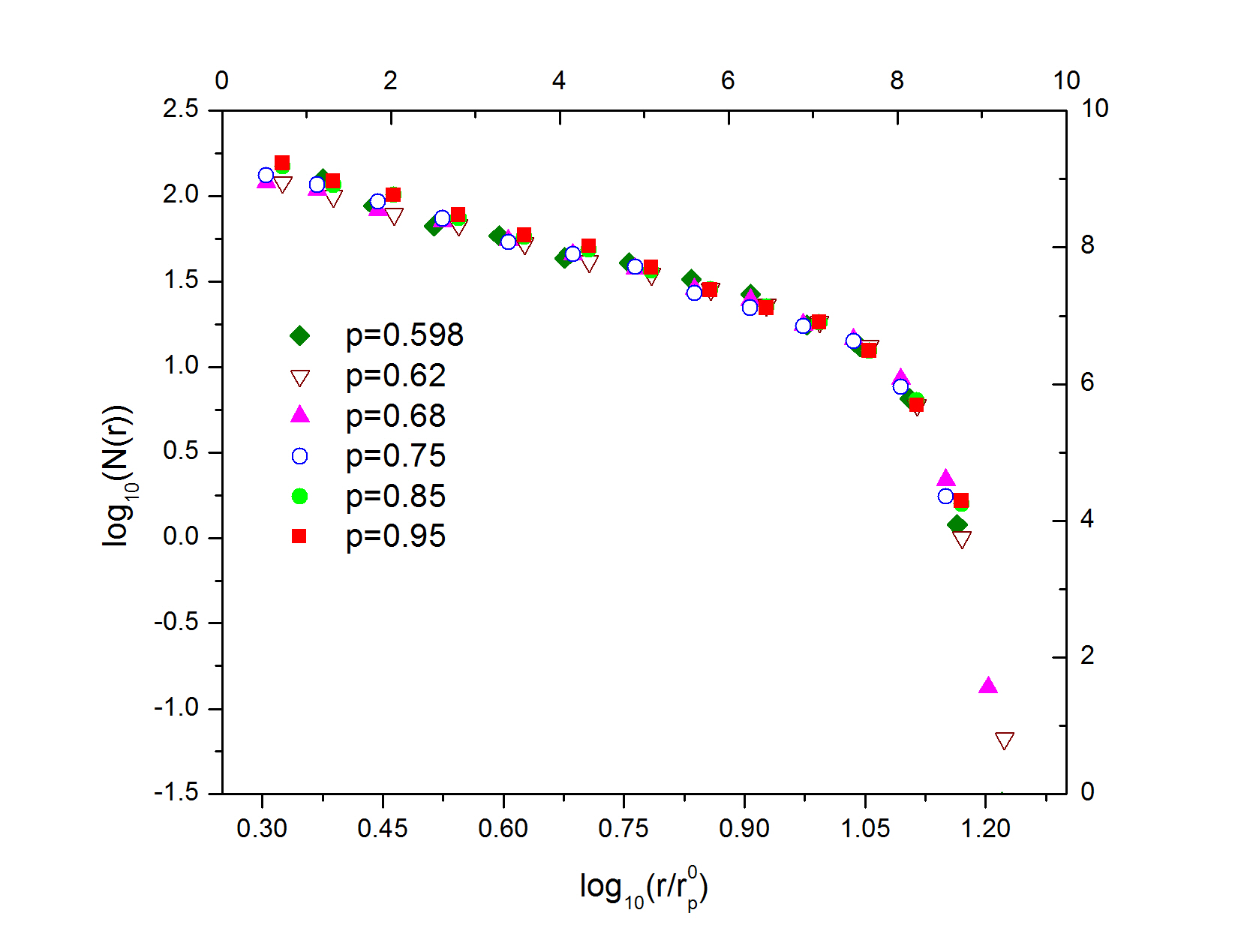}}
\caption{The fitted graph of distribution function of gyration radius.}
\label{gyrationb}
\end{figure}
\\
\\
\textbf{Schramm-Loewner Evolution}\\
In this section we analyze the mentioned curves by means of SLE technique. SLE$_{\kappa}$ is a growth process defined via conformal maps, $g_{t}(z)$, which are solutions of Loewner's equation:
\begin{equation}
\partial_{t}g_{t}(z)=\frac{2}{g_{t}(z)-\xi_{t}}
\label{SLE-eq}
\end{equation}
where the initial condition is $g_{t}(z)=z$  and $\xi_{t}$ (the driving function) is a continuous real valued function which is shown to be proportional to the one dimensional Brownian motion ($\xi_t=\sqrt{\kappa}B_t$) if the curves (assumed to go from origin to the infinity) have two properties: conformal invariance and the domain Markov property\cite{Cardy2}. For fixed $z$, $g_{t}(z)$ is well-defined up to time $\tau_{z}$ for which $g_{\tau_z}(z)=\xi_{t}$. There is a simple relation between $\kappa$  (the diffusivity parameter) and the central charge in conformal filed theory, namely $c=\frac{(6-\kappa)(3\kappa-8)}{2\kappa}$\cite{BauBer}. Therefore $\kappa$ represents the universality class of the model in hand. To extract this parameter, we should follow these steps:\\
- Transform the loops (frontiers of the avalanches) to the curves which go from the origin to the infinity (chordal SLE). To this end we cut loops horizontally and then send its end point to the infinity by the map $\phi(z)=\frac{x_{\infty}z}{z-x_{\infty}}$ in which $x_{\infty}$ is the end point of the cut curve and $z=x+iy$ is the complex coordinate in the upper half plane\cite{BBCF}.\\
- Assume the driving function to be partially constant in each time interval and discretize Eq [\ref{SLE-eq}].\\
- Confirm that $\left\langle\xi_t\right\rangle=0$ and calculate the slope of $\left\langle\xi_t^2\right\rangle-\left\langle\xi_t\right\rangle^2$ versus time $t$ ($\left\langle\right\rangle$ is the ensemble average), i. e. $\left\langle\xi_t^2\right\rangle-\left\langle\xi_t\right\rangle^2=\kappa t$ which yields the diffusivity parameter $\kappa$.\\

For this simulation we have generated the samples of lattice size $200\times 200$ for the Darcy model and $1000\times 1000$ for the SOC model and extracted $\left\langle \xi_t\right\rangle$ and $\left\langle \xi_t^2\right\rangle$. We have found that in two mentioned models $\left\langle \xi_t\right\rangle\simeq 0$. FIG [\ref{SLE}] contains the graph $\frac{1}{\kappa}\left(  \langle  \xi_t^2\rangle-\langle  \xi_t\rangle^2\right) $ versus $t$ for the case $p=p_c$ and $\kappa=3.05 (\pm 0.1)$ for the Darcy model and $\kappa=3.01(\pm 0.1)$ for the SOC model. It confirms the hypothesis that two models lie within the same universality class due to their identical diffusivity parameters. The main point is that this class is the Ising universality class corresponding to $c=\frac{1}{2}$ CFT and $\kappa=3$, i. e. the complex non-linear 2D Darcy model (and also SOC model of water propagation in the porous media) is reduced to the 2D Ising model which is well-understood.

\section{Further Analysis of the SOC Model}\label{num}

The other geometrical quantities relevant to our study are the distribution functions of gyration radius, loop length and the mass of the clusters  ($N(r)$, $N(l)$,$N(m)$ respectively) in SOC model. These quantities behave like $N(r)\sim r^{-\tau_r}$, $N(l)\sim l^{-\tau_l}$, $N(m)\sim m^{-\tau_m}$ up to scales upon which the finite size effect appears. FIG [\ref{gyrationa}] shows the gyration distribution function in which $r^p_0$ is the mentioned scale. This scale is interpreted as the correlation length in the system (the maximum distance from injection point above which the water propagation statistically stops) and can be extracted by fitting the graphs which is done in FIG[\ref{gyrationb}].
\begin{figure}
\centerline{\includegraphics[scale=.40]{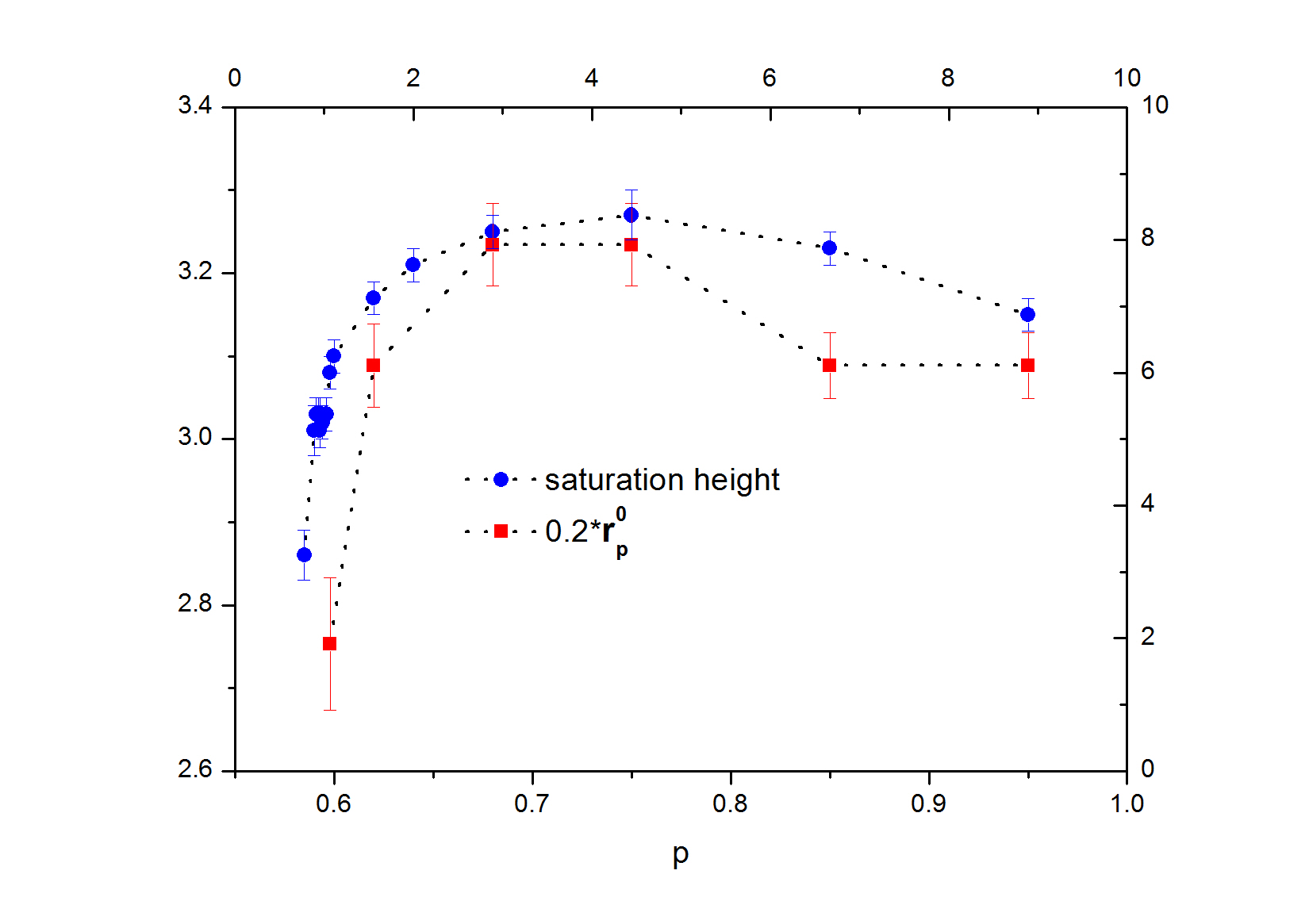}}
\caption{The correlation length $r_0^p$ and the mean amount of water (saturation height $\bar{h_i}$) in the steady state picked nearly around $p=0.68$.}
\label{corr}
\end{figure}
 The $p$ dependence of $r^p_0$ has been represented in FIG [\ref{corr}]. The interesting feature of this graph is that this quantity is maximized around $p=0.68$ in which the average amount of water per site is nearly maximized. This is in contrast to the common belief that the propagation of fluid in the porous systems in $p=1$ is maximum. To inquire this supposition, we have calculated the percolation probability in terms of $p$ which has been shown in FIG [\ref{percolation}]. It can be seen in this figure that the percolation probability is maximum in $p=0.65$ and decreases to the final value $0.0158$ in $p=1$. The tail of this graph can be fitted by the formula $P(p)=P_{1}+a p^{\alpha}e^{\frac{(p-p_0)^2}{\beta}}$ with $P_1=0.0158, a=0.0202, \alpha=0.34,p_0=0.62, \beta=0.0244$. Some rapid oscillations around $p=p_c$ can be seen in this graph, the same as FIG [\ref{FDb}], showing singular behaviors around this point.
\begin{figure}
\centerline{\includegraphics[scale=.40]{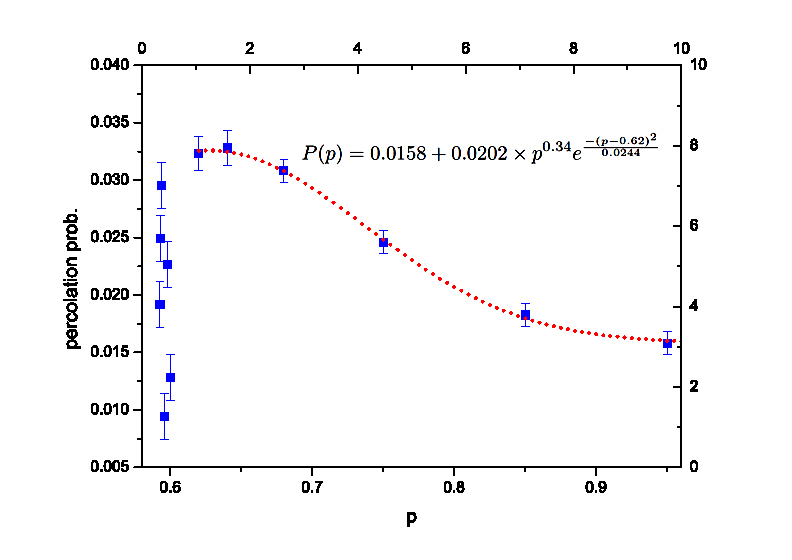}}
\caption{The percolation probability in terms of $p$, picked nearly around $p=0.65$.}
\label{percolation}
\end{figure}
Such singular behaviors at $p=p_c$ is seen in the exponents of the distribution functions, i. e. $\tau_r, \tau_l, \tau_m$ as indicated in FIG [\ref{taua}]. After rapid oscillations, which shows their singular behaviors, these quantities tend to a final value at $p=1$ which are $\tau^{p=1}_l=1.28,\tau^{p=1}_m=1.16,\tau^{p=1}_r=1.43$. We also have calculated the distribution function of the number of topplings in an avalanche, $N(n_t)\sim n_t^{-\tau_{n_t}}$ which has been shown in FIG [\ref{taub}].\\
\begin{figure}
\centerline{\includegraphics[scale=.40]{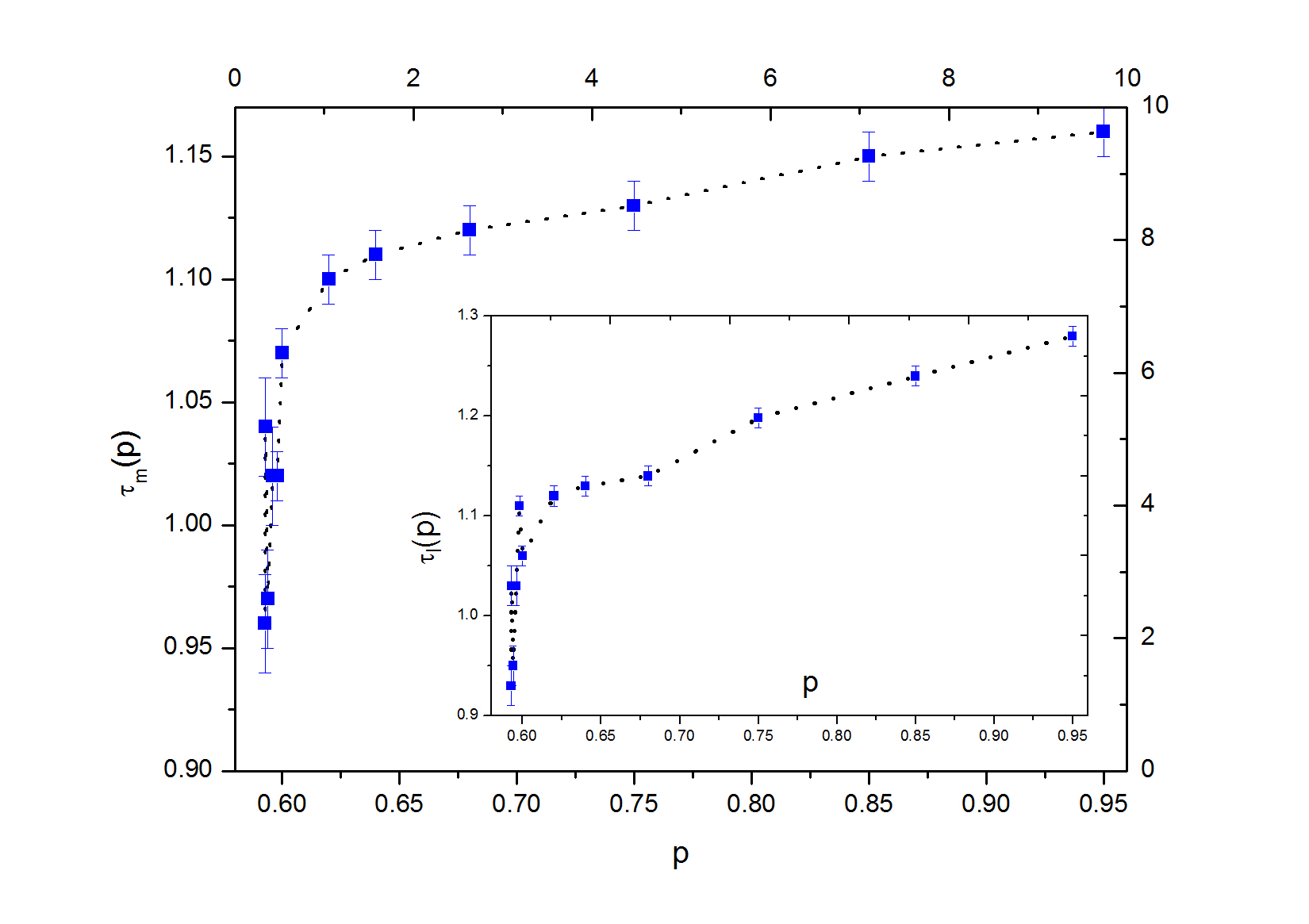}}
\caption{$\tau_m$ and $\tau_l$ versus $p$ defined in text.}
\label{taua}
\end{figure}
\begin{figure}
\centerline{\includegraphics[scale=.40]{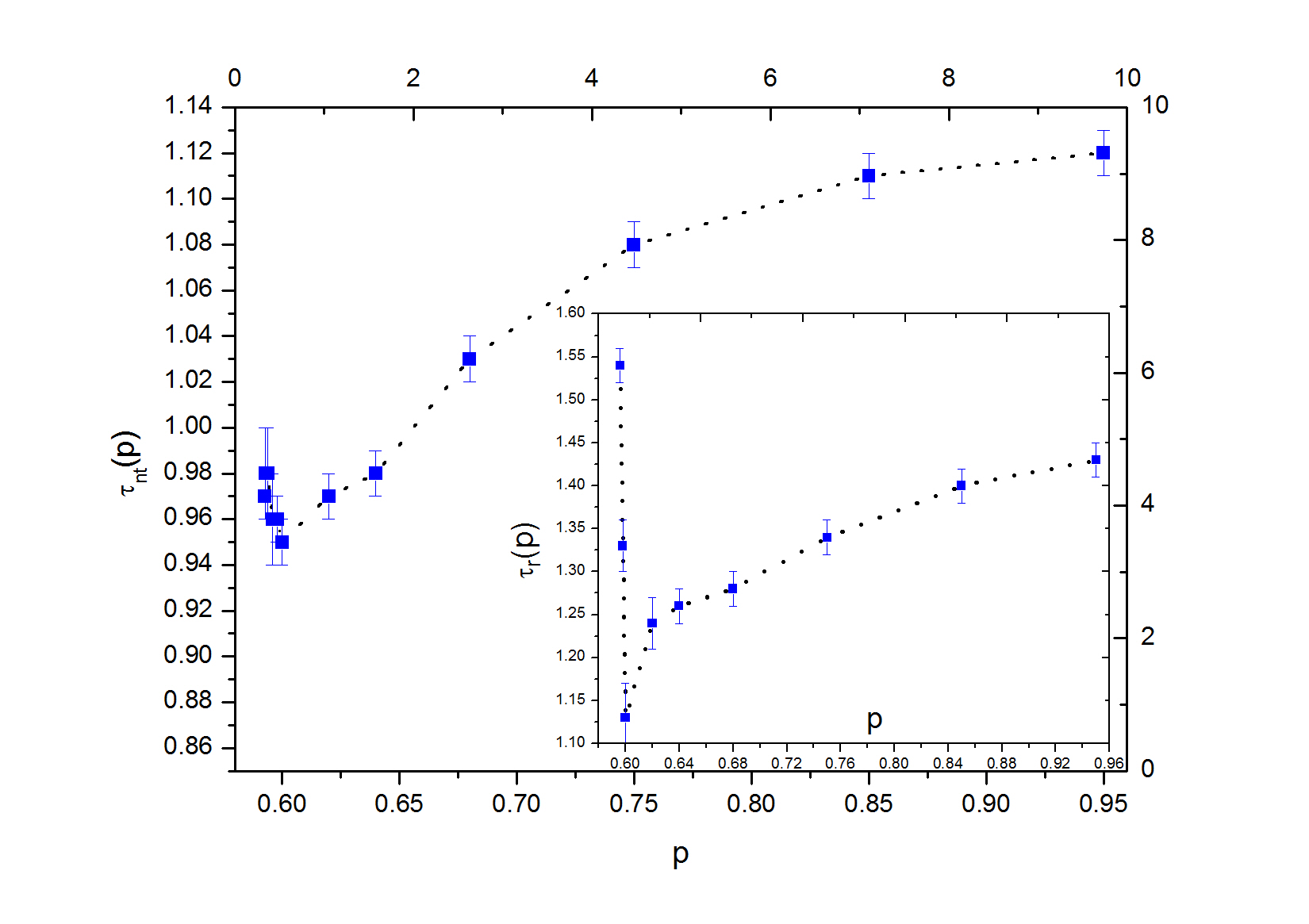}}
\caption{$\tau_{n_t}$ and $\tau_r$ versus $p$ defined in text.}
\label{taub}
\end{figure}
\begin{figure}
\centerline{\includegraphics[scale=.40]{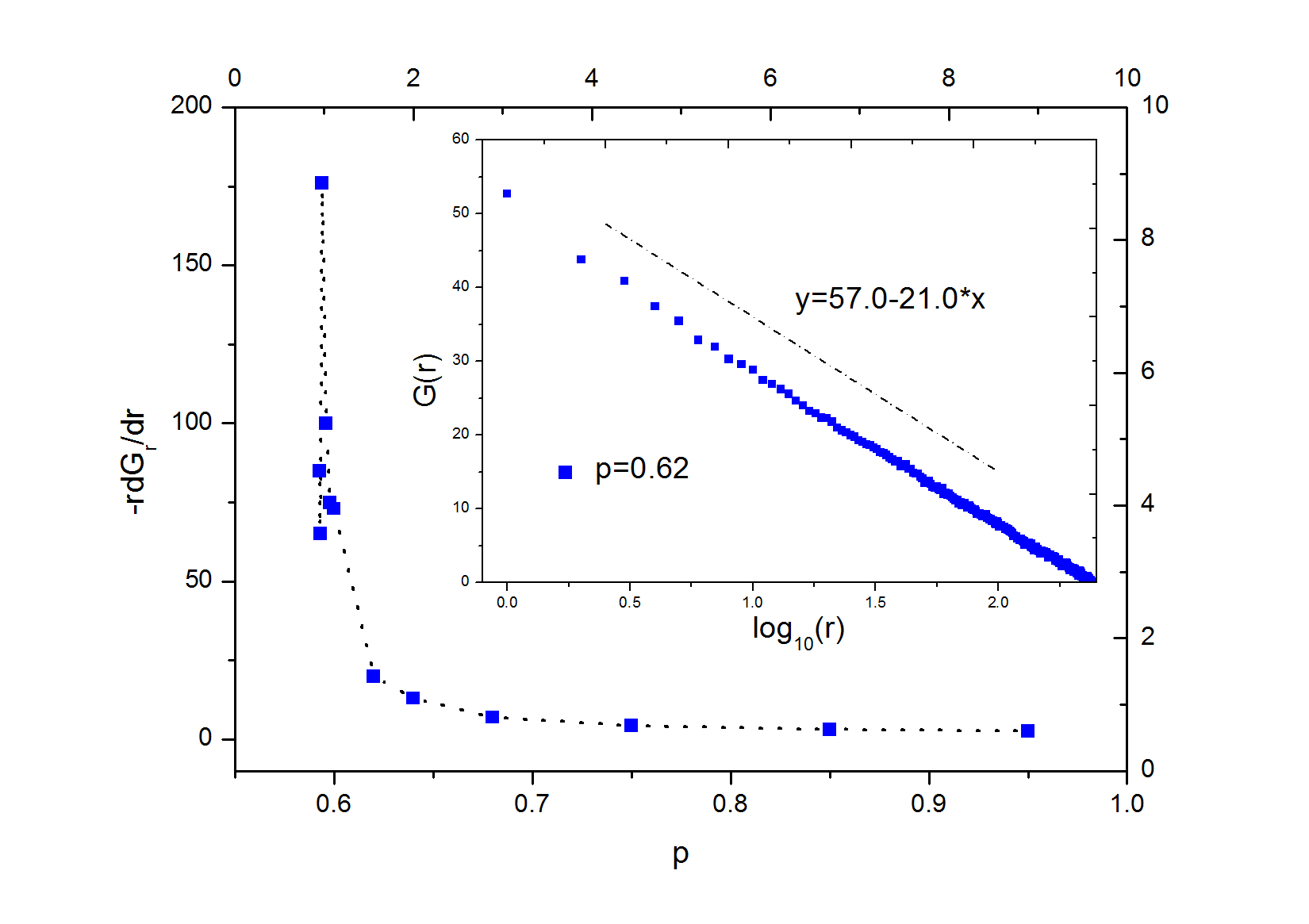}}
\caption{The Green function ans its slope defined in text.}
\label{Greena}
\end{figure}
\begin{figure}
\centerline{\includegraphics[scale=.40]{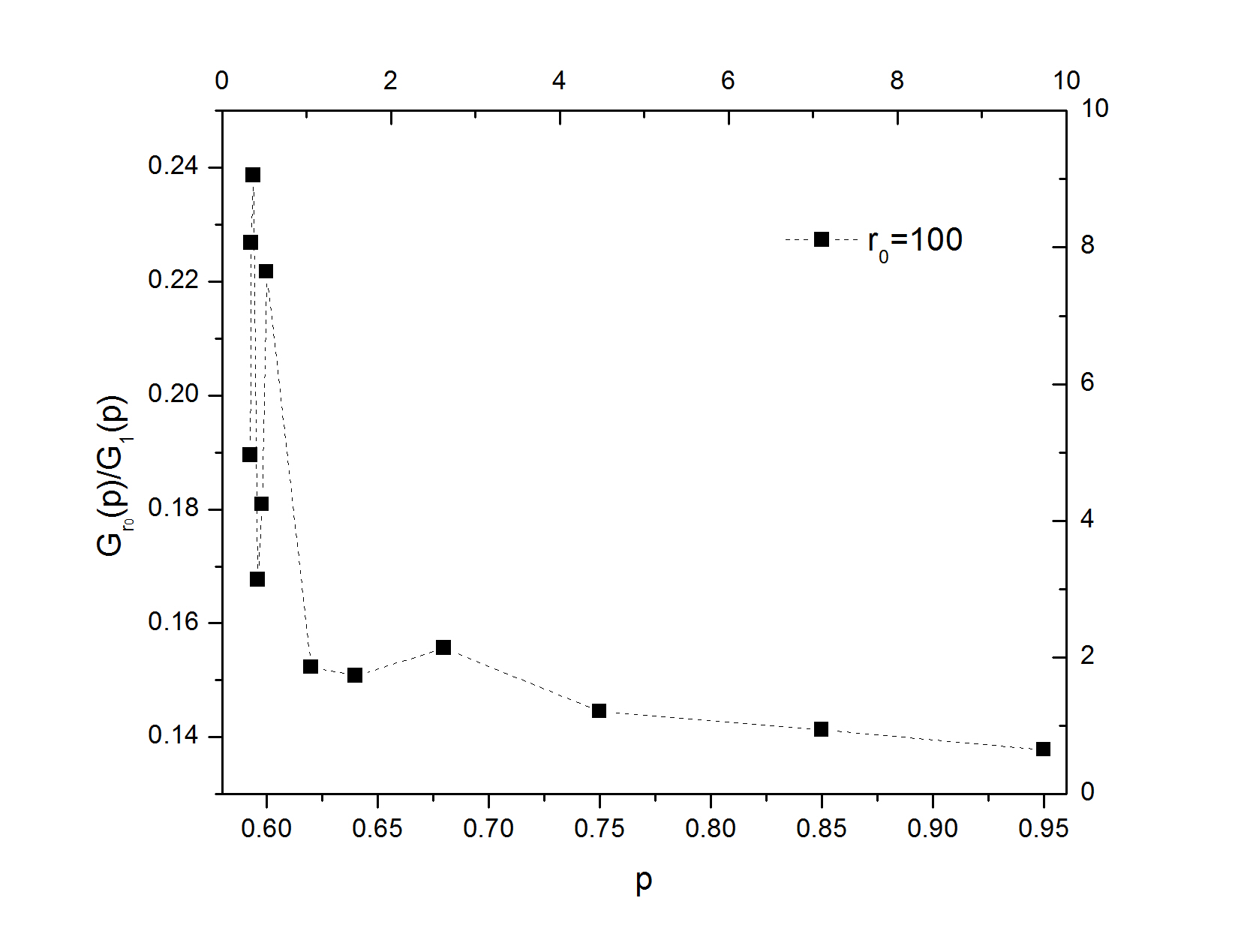}}
\caption{$\frac{G_p(r=100)}{G_p(r=1)}$ versus $p$.}
\label{Greenb}
\end{figure}
\\
\textbf{The Green function} $G(\vert i-j\vert)$ is defined as the number of topplings occurring in the site $j$ (up to a normalization factor) if one add a grain to the site $i$. We found that the Green function is logarithmic for all $p\geq p_c$ just like the 2D BTW in which the Green function is the inverse of the matrix $\Delta$ and is a logarithmic function. Representing $G(r)=a_p-b_p\log(r)$, $b_p=-r\frac{dG}{dr}$ has been shown in FIG [\ref{Greena}] in which again the singular behavior is apparent. Let's define $f_{r_0}(p)\equiv\frac{G_p(r_0)}{G_p(r=1)}$ as the function which shows the number of topplings (in a typical point $r_0$) as a function of $p$. This function has been sketched in FIG [\ref{Greenb}]. Two behaviors are apparent; singular behavior around $p_c$ and a local maximum at $p\simeq 0.68$ which is the point at which the saturation height ($\bar{h_i}$ in the steady state), $r_0^p$ and the percolation probability are maximum.

\section{Conclusion}
     In this paper we have analyzed the water propagation in the 2D porous medium. To this end we have investigated the well-known Darcy model and also introduced a self organized critical model corresponding to this dynamics. Using Schramm-Loewner evolution (SLE) technique, we have numerically shown that, at $p=p_c$, which is $0.5927$ in the square lattice, these models belong to the Ising universality class. Further analysis on the SOC model demonstrated that the percolation probability is maximum around $p\simeq 0.68$.

\end{document}